\documentclass[modern]{aastex61}

\usepackage{tabulary}

\providecommand\citet{\cite}
\providecommand\citep{\cite}

\usepackage{url}
\usepackage{hyperref}
\hypersetup{colorlinks=false,pdfborder={0 0 0}}

\shorttitle{Research source code availability \& link persistence}
\shortauthors{}

\begin{document}

\title{Schroedinger's code: A preliminary study on research source code availability and link persistence in astrophysics}

\author[0000-0003-3477-2845]{Alice Allen}
\affiliation{Astrophysics Source Code Library, ascl.net}
\affiliation{University of Maryland, Department of Astronomy, PSC 1208, 4254 Stadium Drive, College Park, MD 20742
}

\author[0000-0003-1774-3436]{Peter J. Teuben}
\affiliation{University of Maryland, Department of Astronomy, PSC 1208, 4254 Stadium Drive, College Park, MD 20742
}

\author{P. Wesley Ryan}
\affiliation{Astrophysics Source Code Library, ascl.net}

\correspondingauthor{Alice Allen}
\email{aallen@ascl.net}

\begin{abstract}
We examined software usage in a sample set of astrophysics research articles published in 2015 and searched for source code for the software mentioned in these research papers. We categorized the software to indicate whether source code is available for download and whether there are restrictions to accessing it, and if source code is not available, whether some other form of the software, such as a binary, is. We also extracted hyperlinks from one journal's 2015 research articles, as links in articles can serve as an acknowledgment of software use and lead to data used in the research, and tested them to determine which of these URLs are still accessible. For our sample of 715 software instances in the 166 articles we examined, we were able to categorize 418 records as to availability of source code and found that 285 unique codes were used, 58\% of which offer source code available online for download. Of the 2,558 hyperlinks extracted from 1,669 research articles, at best, 90\% of them were available over our testing period.
\end{abstract}%

\section{Introduction} \label{sec:intro} 
 
\begin{quote}
``Is a paper for which the author feels that it is not worth making the code available worth reviewing?" \citep{VitekKalibera2011}
\end{quote}

Astrophysics is, like nearly every discipline \citep{Ludaescheretal1016,Howisonetal2015,hettrick2014}, dependent on software for many aspects of the research process \citep{MomchevaTollerud2015}, from gathering, reducing, and cleaning data, to modeling and visualizing data and creating analyses of enormous amounts of data \citep{Soito_2016,Brownetal2007}. Reliance on software cannot be overstated, yet as computational methods have proliferated, research has become less transparent, reproducible, and falsifiable because these methods have not been shared or made available along with the research they enable \citep{Baker2016,Gobleetal2016,Stoddenetal2016,Marwick2015,Shamiretal2013,Morinetal2012}. 

Certainly as these methods started to be used, it was inconvenient to share long or complex codes if they could not be included within research papers; sharing software was typically limited to those within close physical proximity of the software author. As technology has advanced, sharing source code with researchers everywhere is possible; still, practices have not caught up with what technology now enables. Reasons for this are varied and have been covered in the literature \citep{Inceetal2012,barnes2010,Stodden2010,Weineretal2009}, in sessions at astronomy meetings \citep{Teubenetal2014,2013ASPC..475..383A} and more broadly at multidisciplinary workshops \citep{Gobleetal2016,Ahaltetal2015,Katzetal2014}. 

We will not discuss the reasons here except to acknowledge they exist, and that the literature, meeting and workshop reports that cover barriers and reluctance to sharing also include calls for greater openness and availability of source code. Indeed, some have called for peer review of research software \citep{Joppaetal2013} and have worked to implement it \citep{Smithetal2017,limare2012}. Organizations and events dedicated to improving scholarly communication and free access to research artifacts are actively working to increase transparency and reproducibility (Force 11 \citep{softwarecitationprinciples}, OpenCon\footnote{\url{http://www.opencon2017.org/}}, Workshop on Sustainable Software for Science: Practice and Experiences\footnote{\url{http://wssspe.researchcomputing.org.uk/}}, Software Sustainability Institute\footnote{\url{https://www.software.ac.uk/}}) and government agencies are requiring that research products paid for by taxpayers, particularly data and software, be available to them \citep{nasadatacitation}. Journal practices have evolved and may request, or even require, that source code and data be made available, as is seen with \textit{\href{http://www.nature.com/authors/policies/availability.html}{Nature}} and \textit{\href{http://www.sciencemag.org/authors/science-editorial-policies}{Science}}, and Elsevier has declared a new ``academic content class" called \href{https://www.elsevier.com/authors/author-services/research-elements/software-articles/original-software-publications}{Original Software Publications}, which requires peer review of the software in that content class. Domain and general journals focused on peer review of software also exist, such as \textit{Source Code for Biology and Medicine\footnote{\url{https://scfbm.biomedcentral.com/}}}, \textit{Image Processing On Line\footnote{url{http://www.ipol.im/}}}, and \textit{Journal of Open Source Software\footnote{\url{http://joss.theoj.org/}}}.

Infrastructure and services such as the NASA Astrophysics Data System\footnote{\url{http://adsabs.harvard.edu/}} (ADS), GitHub\footnote{\url{https://github.com/}}, the Astrophysics Source Code Library\footnote{\url{http://ascl.net/}} (ASCL), Zenodo\footnote{\url{https://zenodo.org/}}, and Figshare\footnote{\url{https://figshare.com/}} make it easier to share and discover useful research software. With more ways to make these programs available for examination, we wonder: How many codes that are used and clearly identified in research can be examined by other researchers? And how much research is enabled by software without it being made explicitly clear in an article that a computational method was used? These are questions we are starting to research.

We are not the first to consider these questions. A wide-ranging project undertaken by \citet{CollbergProebsting2016} included studying papers by Computer Systems researchers to find those with results dependent on software and to get the source code if possible. Among other questions, they also studied whether the software would build within 30 minutes, or at all, without help from the software authors, whether National Science Foundation (NSF) funding has an impact on sharing source code, and examined results from other studies on research software availability. In their own study, out of the articles they examined that were backed by software, ``for 56.2\% of them the authors were able to make that [source] code available."

In a robust study of biology articles, \citet{HowisonBullard2016} examined software citations and accessibility, and found that ``software is frequently inaccessible" and that ``only between 24\%-40\% provide source code." Their study covered whether code could be accessed, whether software was licensed in a way to permit reuse, how well authors provided software creators with credit for these scientific contributions, and also whether version information was provided in the articles.

Here we report on our preliminary efforts on the first question, which we undertook to find out how we might do a comprehensive study to answer this and related questions. The availability of hyperlinked resources (especially code and data) related to published research is also important for the reproducibility, reliability, and falsifiability of scientific results, so we also examined the extent to which hyperlinked resources are accessible.

Unlike \citet{HowisonBullard2016} and \citet{CollbergProebsting2016}, we restricted our study to the issue of transparency of the computational methods  -- can the reader see what was done? -- which certainly includes the ability to identify the software and find it, but we did not tackle the other clearly important issues of how well codes were cited, how they were licensed, whether they would build, nor whether the version used in an article was still available in this preliminary study. We also sought to determine the persistence of hyperlinks included in research papers, as such links are often for software and data needed to ensure the transparency of the research.

\section{Issues in answering the research questions} \label{sec:style}

Several issues arise in trying to conduct this type of study. Research articles do not always include what software was used; some do not even acknowledge that software was used, yet some tasks described are too cumbersome to not have had some sort of software involved. Software often unintentionally gets left out, as in this text:
\begin{quote}
With the observed IR PSD, we simulate light curves using the method of Timmer \& Koenig (1995). Gaps identical to the IR light curve were artificially introduced in the simulated light curves and the resulting PSD is compared to the PSD of the original light curve. In the second test, we introduce gaps in the RXTE X-ray light curve (which does not have any gaps) identical to the simultaneous IR light curve, fill these gaps using similar technique as in the IR data, and compare the PSDs. \citep{Kalamkaretal2016}
\end{quote}

``... fill these gaps using similar technique as in the IR data..." In other words, a code was used (pyLCSIM, \cite{Campana2017}), though it is not mentioned by name. Maybe it is obvious to others that some unnamed code was referred to in that phrase, but it was not obvious to the first author when she first ran across this text.

It is easy to determine a paper used research software when the text says, ``The photometric reduction was performed with AstPhot, a synthetic aperture photometry tool" \citep{Tubianaetal2015}, or, ``In our calculations we made use of the public code developed by Dany Page, NSCool" \citep{BonannoUrpin2015}. 

Or even when the software used is not named, as in \citet{Santos-Sanzetal2015}, which said ``A common reduction software programmed in IDL was used for the photometry of all the previously calibrated images." We do not know what software was used, but we do know something was.

One of our longer term motivations for doing this work is to refine our ability to search for code use programmatically, and as we continue to develop the data we have gathered, we will use it to further train and benchmark our automated method. One of the difficulties in searching, however, is when the software is not mentioned by name in text, such as ``Original light curves were processed and corrected from phenomenological effects such as outliers, jump and drifts according to the method of Garc\'ia et al. (2011)." \citep{Vrardetal2015} or ``We performed ... data reductions following the procedures outlined in Taddia et al. (2013b)." \citep{Taddiaetal2015}. 

Though we have selected specific examples to demonstrate some of the challenges in trying to determine the availability of software used in research, we certainly do not mean any disrespect to any author (of article or software) or journal. These examples follow norms in research publishing (clearly there is nothing wrong with referring the reader elsewhere for background or more information!) but still present a challenge to finding computational methods. We want to shed a light on one aspect of science -- software use -- that has an impact on the transparency and replicability of astrophysics, and we do not fault any author for following standard practices and discipline norms.

\section{Methods} \label{sec:methods}

\subsection{Planning}
We are considering examining software use in papers over time. Though the Astrophysics Source Code Library (ASCL) was founded in 1999, it started being indexed by the NASA Astrophysics Data System (ADS) only in January 2012. One impetus for conducting such research is to see what impact the ASCL and the other efforts to improve source code availability mentioned in the Introduction may have on the community. We think five-year intervals are broad enough to show changes in source code accessibility without being too onerous to do, and selected 2010 as a good start for the cycle of five-year intervals for several reasons. First, it's a nice round number, but more importantly, it would give us data on software availability before the indexing of the ASCL by ADS began; this is also before many other efforts and organizations with similar concerns were started or could have an impact, including Force11, Workshop on Sustainable Software for Science: Practice and Experiences, and the Software Sustainability Institute, and also before NASA and other funding agencies required data (including software) management plans.\footnote{\url{https://www.nasa.gov/open/researchaccess/frequently-asked-questions\#dmpfaqs}} 

We searched ADS Bumblebee to learn how many papers each of three journals, the \textit{Astrophysical Journal}, \textit{Astronomy and Astrophysics}, and \textit{Monthly Notices of the Royal Astronomical Society} (\textit{ApJ}, \textit{A\&A}, and \textit{MNRAS} respectively), published in 2010 and 2015 (see Table \ref{tab:adssearch}). 

\begin{deluxetable}
{c|cc}
\tablecaption{ADS Bumblebee searches \\ 
Total papers for the given years\tablenotemark{a} \label{tab:adssearch}}
\tablehead{
\colhead{Journal} & \colhead{2015} & \colhead{2010}
}
\startdata
\textit{ApJ} & 3,055 & 2,539 \\
\textit{MNRAS} & 3,140 & 1,987 \\
\textit{A\&A} & 1,805 & 1,932 \\
\enddata
\tablenotetext{a}{Information retrieved 2017 June 4 }
\end{deluxetable}

\textit{ApJ}, \textit{MNRAS}, and \textit{A\&A} collectively published 8,000 papers in 2015 and 6,458 papers in 2010. We eventually want to look at software use in 2010, but for this preliminary study, we chose to start our research with papers published in 2015 to gather information on recent research software availability. Using articles from roughly the same time period as studies in other disciplines makes it feasible to compare our results with those other studies and other researchers may find this study useful for that reason.

Though our plans are to extend this research to include \textit{ApJ} and \textit{MNRAS} articles, each of these presented special concerns for our initial foray into answering our research questions. In 2015, \textit{MNRAS} had one editor in particular who was very active in promoting software citation and open access to code; additionally, these papers are still behind a paywall. \textit{ApJ}, though offering free access to 2015 articles, did not officially allow publication of software papers nor formal software citation until 2016, though, wisely, these rules were not rigorously enforced and software papers and citations for codes certainly appear in its 2015 issues. \textit{A\&A} fell between these two; in 2015, it published software papers and allowed software citation in an ADS-capturable way but (so far as we know) did not have an editor who was really pushing to improve the journal in this way as \textit{MNRAS} did. We felt papers from \textit{A\&A} would offer a balanced picture. Also, the journal offered a reasonable rather than overwhelming number of articles and it was relatively easy to download the year's worth of papers to examine. The papers are free access and readily available to anyone who would like to check our research (which we encourage). We therefore conducted this preliminary study using 2015 \textit{A\&A} articles.

\subsection{Article selection}
There were 1,805 articles published in \textit{A\&A} in 2015. We removed the 131 Letters from consideration. This left 1,653 research articles and 21 Errata. We selected 10\% of these remaining articles for examination by ordering the papers by article number (an identifier given by the publisher) and selecting every 10th paper, giving us a sample set of 166 papers to examine. 

The distribution of selected papers by publication month varied from a low of nine for October (out of 127 articles that month) to a high of 20 for November (out of 143 articles that month). November's issue included a special focus on Rosetta; perhaps focus articles had a firm due date, resulting in an abundance of consecutively-numbered articles for that issue. 

Three errata were among the selected articles; we replaced these with the original papers in two cases, as the original articles were published in 2015; the remaining erratum was for a paper from 2014 so we replaced the erratum with the paper immediately after it in the ordered list. Although anyone following the procedures above should select the same articles we used, a table of article IDs and links to the PDFs of those we chose is in Appendix \ref{app:C}.

\subsection{Looking for code in all collected papers}
We skimmed each article from start to end for evidence of software use. As \citet{HowisonBullard2016} did, we looked not just for software indicated by a formal citation, but also for software ``mentions", as they put it, to include codes in text, figures, footnotes, and appendices. Unlike Howison \& Bullard, we also captured text that indicated software was very likely used though was not specified. 

We considered what to include or perhaps exclude. Our focus is on source code; we weren't sure initially whether we wanted to find software in all its incarnations -- binaries, web calculating services, commercial code mentions, software written in the 1980s and before -- and then sort through whatever we find, or gather just source code. In the end, we excluded two large Government-run software systems (NOAA's Global Forecast System and the Comprehensive Large-Array-Data Stewardship System) and computational methods that are essentially ``astro/scientific arithmetic" (as we came to think of them) or manipulations, such as Fourier transform calculations, but otherwise included everything else. We looked specifically for mentioned use of software; if an author thought a script was important enough to mention, we counted it. We admit that where to draw the line on what to include is somewhat arbitrary and may depend on one's specific area of research, as what is common in one particular domain may not be used in another, and that others doing similar research may draw the line in a different place.

After our initial read, we did a full-text search of each of the 166 article PDFs for a minimum set of search terms, listed in Table \ref{tab:searchterms}, to try to ensure no computational methods were overlooked.

\begin{deluxetable}
{ccccccc}
\tablecaption{Search terms used to find code methods \label{tab:searchterms}}
\tablehead{}
\startdata
code & python & numeric & routine & script & fortran & program \\
http & model & software & librar & algorithm & IDL & ftp \\
pipeline & github & procedure & iraf & package & comput & recipe \\
\enddata
\end{deluxetable}

Additional search terms were used occasionally for specific papers when the read-through of an article suggested such searching might be useful. 

We copied text that indicated definite or possible use of computational methods from the article to create a datafile with 166 records, one for each paper. 

As a quality check for the search method, eight articles, 4.8\% of our sample chosen for this preliminary work, were examined by others (members of ASCL's Advisory Committee) and their results compared against the first author's. 

We went through the collected text to organize and atomize the information. The 166 papers yielded a dataset of 715 records, with each record representing possible use of one computational method.

In a first pass of these data, we identified software registered in the ASCL, though we did not catch all such entries. In our second pass, we looked to see which records identified software by name or had a URL or other very definite indication for software that was readily searchable. Subsequent passes resulted in increasingly cleaner data and deletion of entries found to be not software related. Of our 715 records, 418 of them were ultimately used for the research discussed in the rest of this article.

The records we did not include in this study do not offer readily searchable information for software; some, for example, refer the reader to another paper or papers or may state that a code method was used but offer no other information about it. Other records are for what we refer to as ``hidden software," which is where article text strongly suggests that a computational method had to have been used, but does not provide any information whatsoever as to what the method was. We are interested in trying to determine how much research software is hidden so collected these data in addition to information that is more transparent about software use; we hope to use these entries in future research but they are outside the scope of this article so are not considered here.

We looked for download sites for the software in the 418 records by doing the following:

\begin{enumerate}
\item Check to see whether the code had an ASCL entry
\item Look in the examined paper for URL or other location information; if the code had a citation, in the paper cited for a URL or other location information
\item Use Google to search for the name of the code, its author(s), and other relevant information
\item Look on the author's and/or project website
\end{enumerate}

We took an additional step for some software; we sent one of the ASCL's standard emails asking whether there was a download site for the source code to the software author(s), and in one case, one article author, with the intent of registering the software in the ASCL if the code was available. 

\subsection{Categorizing availability}
Once we had completed our search for a download site for a code, we categorized the code's availability with one of the designations shown in Table \ref{tab:categories}.

\begin{table}[]
\centering
\caption{Category explanations}
\label{tab:categories}
\begin{tabular}{p{2cm}p{11cm}}
Category & Explanation                                                                                                                                                                                                                                                                                                                                                            \\
A & We found source code available for download without reservation or impediment.                                                                                                                                                                                                                                                                                         \\
B & We found an executable file, such as a binary or other compiled file, available for download without reservation or impediment.                                                                                                                                                                                                                                        \\
C  & We learned from correspondence, website, or other authority that the source code is available only to collaborators. \\
N & We did not find a way to download the source code. This could be for several reasons, such as no download site exists, one exists but we were unsuccessful in finding it, or a site for the code exists but the URL for the download did not work when we tried it. \\
W & The code can be used as an online service, but we could not find a download site for the source code. In cases where the software is available for use as an online service and the source code is available for download without reservation or impediment, the code was categorized as A; if available as a binary and a web service, the code was categorized as B. \\
GS & The source code is available for download behind a soft gate, which means one must provide information of some kind online to get immediate automated access to download the code.\\
GH & The source code is stated to be available and is behind a hard gate that requires human action of some kind to receive the source code, such as an email to the author, an online form without immediate automated access, or a requirement to attend training before gaining access to the source code.\\
P1 & The source code itself is available for purchase (such as \textit{Numerical Recipes} codes).\\
P2 & The software is commercial software purchased as an executable (such as Feko, Mathematica, and IDL).\\
O  & Anything not fitting a designation listed above.                                                                                                                                                                                                                                                                                                                      
\end{tabular}
\end{table}

Some software packages are contained within a larger package, such as NuSTARDAS in HEASoft \citep{HEASOFT}, and were attributed to the larger package and categorized under that larger package. In other words, if a paper stated it used NuSTARDAS, the software is listed as HEASoft for categorization. There are exceptions to this, such as DAOPHOT \citep{DAOPHOT}, which was attributed to IRAF \citep{IRAF1,IRAF2} when a paper stated it had used IRAF, but is listed separately when it appeared alone. We recognize that it is somewhat unfair to some packages to do this, but as we are primarily determining code availability, not specific package, routine, or function use, have made this pragmatic and in practice occasionally arbitrary decision.

We initially categorized all records for codes in the ASCL as \textit{A}, and then tested the availability of these packages by verifying that the links the ASCL has for these codes worked, and where they did not, found a new download link, or failing to find one, adjusted the categorization as needed. 

Some entries needed close examination, such as when a link was retrieved that went to a site that offered both data and software. If only the data from the site were used in the paper under examination, we did not include it in our analysis. In other words, only when the authors of a selected paper used a code on a site did we include it. One such case is MILES; though software on that site generated available data, a researcher may use only the generated data and not run any code from that site herself.

Algorithms obviously can be implemented in different languages and with different coding techniques. The Lomb-Scargle periodigram algorithm \citep{Lomb1976,Scargle1982} is widely used and multiple implementations of it exist\footnote{see \citet{VanderPlas2017} for a discussion of various implementations}; not all articles using it specified which implementation was used. We scored this as \textit{P1} when a \textit{Numerical Recipes}  \citep{Press1986,Press1992C,Press1992F,Press2002C} implementation was specified, \textit{W} when an online service was used, and the unknown cases as \textit{N}. We struggled with this a bit, but as said above, we are determining code availability, and if we do not know the implementation used, we cannot examine that specific software. 

We emailed some authors when we did not find a download site for a code through other methods using the ASCL's software request email template (Appendix \ref{app:B}); this is a standard practice for the ASCL editors when we are unable to find a download site for software we come across in papers, as we hope to register the code in the ASCL. We sent 46 emails seeking information on code availability in our data file for 49 unique codes and received 19 replies. 

We started registering software that meets ASCL criteria as we found it in working through the records, but had to abandon doing so for the duration of our research period due to time and will add these codes at a later date. 

\subsection{Checking links in papers}
As mentioned in the Introduction, hyperlinks in articles may point to software or data, informing the user to the location of resources used in the journaled research, the loss of which may make the research less transparent. We examined whether the hyperlinked resources were accessible by writing a set of scripts, accessible at https://github.com/teuben/ascl-tools, to automate the gathering and testing of hyperlinks in research papers.

The script \textit{process\_pdfs.py} extracts every hyperlink from the PDFs of every research paper published in \textit{A\&A} in 2015 (excluding Letters) using pyPdf v1.13 \citep{PYPDF}, other than those hyperlinks that point to eight common and reliable sites (aanda.org, linker.aanda.org, arxiv.org, ascl.net, dx.doi.org, dexter.edpsciences.org, adsabs.harvard.edu, ui.adsabs.harvard,edu), those that use the doi:, email:, or mailto: protocols, and those that contain an at (@) sign, and stores them in the SQLite database links.sqlite. (All hyperlinks containing an at sign in our set of papers were email addresses with no protocol specified.)

The script \textit{check\_links.py} tests each HTTP(S) link and stores the response code and message returned by the server, or, in the case of an error that prohibits the sending of a request to the link (e.g. a malformed URL), a custom error code represented by a negative number. Links with no protocol specified are assumed to be HTTP. 

After the first run of this script, we edited the database to correct those malformed links that could be corrected (e.g. http:\textbackslash{}\textbackslash{} instead of http://) and remove those that could not be (e.g. a spurious `a'). We debated whether we should be doing even this minimum editing, but decided for it, as it is reasonable to assume a reader would know how to correct these small errors. 

We ran this script four times on four dates from two different locations on our database of hyperlinks. The database generated is available at links.sqlite in the aforementioned GitHub repository.  We checked FTP links by hand four times by entering them manually into our usual web browsers, Opera for one of us and Firefox for the other.

Our script \textit{link\_data.py} gathers the statistics from the database that we use in this paper: how many links consistently worked or failed to work, how many links inconsistently worked or failed for inconsistent reasons, and a count of links that consistently returned each attested error code.

\section{Results} \label{sec:results}
 
\subsection{Software in articles} \label{subsec:resultssoftware}
The 166 papers we examined generated 715 instances of definite or possible software use; of these, we have examined 418 (58.5\%) of the entries, those with codes identified by name, URL, ASCL ID, or otherwise easily determined to be software, to learn how available these packages or routines are. The other 41.5\% of the entries will be examined further at a later date for possible additional research. 

We categorized the 418 records examined into one of our ten categories; the number and percentage of entries in each category are shown in Table \ref{tab:catsummary}.

\begin{deluxetable}
{c|cc}
\tablecaption{Software availability categorization summary \label{tab:catsummary}}
\tablehead{
\colhead{Category} & \colhead{Number of} & \colhead{Percentage} \\
\colhead{} & \colhead{entries} & \colhead{of entries}\\
}
\startdata
A & 262 & 62.7\% \\
B & 26 & 6.2\% \\
C & 4 & 1.0\% \\
N & 70 & 16.7\% \\
W & 21 & 5.0\% \\
GS & 5 & 1.2\% \\
GH & 16 & 3.8\% \\
P1 & 6 & 1.4\% \\
P2 & 6 & 1.4\% \\
O & 2 & 0.5\% \\
Total & 418 & 100\% \\
\enddata
\end{deluxetable}

The 418 software usage instances are for 285 unique codes; those used in four or more papers are shown in Table \ref{tab:shortfreq}; a frequency table for all 285 unique codes is provided as Appendix \ref{app:A}. Of the 285 codes used in the 166 papers, 229 were used in only one paper if we count use of the Lomb-Scargle periodigram algorithm separately, which we do because of its different implementations. 

\begin{deluxetable}
{c|cc}
\tablecaption{Unique code category and frequency (4 or greater) \label{tab:shortfreq}}
\tablehead{
\colhead{Code name} & \colhead{Category} & \colhead{Frequency} \\
}
\startdata
IRAF & 
A & 
31 \\
SExtractor & 
A & 
10 \\
HIPE & 
B & 
7 \\
GILDAS & 
A & 
5 \\
BGM & 
W & 
4 \\
CASA & 
A & 
4 \\
CIAO & 
A & 
4 \\
DAOPHOT & 
A & 
4 \\
DAOSPEC & 
GH & 
4 \\
MOOG & 
A & 
4 \\
MPFIT & 
A & 
4 \\
SAS: Science Analysis Software & 
GS & 
4 \\
XSPEC & 
A & 
4 \\
\enddata 
\end{deluxetable}

The number of unique codes used in the examined papers and the categories we assigned them to are in Table \ref{tab:uniqcodecats}.

\begin{deluxetable}
{c|cc}
\tablecaption{Unique code category summary \label{tab:uniqcodecats}}
\tablehead{
\colhead{Category} & \colhead{Number of} & \colhead{Percentage of unique} \\
\colhead{} & \colhead{unique codes} & \colhead{codes in category} \\
}
\startdata
A & 162 & 56.8\% \\
B & 14 & 4.9\% \\
C & 4 & 1.4\% \\
N & 63 & 22.1\% \\
W & 16 & 5.6\% \\
GS & 2 & 0.7\% \\
GH & 10 & 3.5\% \\
P1 & 6 & 2.1\% \\
P2 & 6 & 2.1\% \\
O & 2 & 0.7\% \\
Total number of unique codes & 285 & 100.0\%  \\
\enddata
\end{deluxetable}

Most software packages were used in only one paper in the sample. The question arose as to whether this software was ``use once and never again." The answer is overwhelmingly no. We did full text searches (using ADS's Bumblebee) on 20\% of these code names and found previous or subsequent use for almost every one of them. The median of the results from ADS was sixty-five articles. Though it is likely some results retrieved by doing these full text searches may not truly indicate use of the software, we are confident these results do show that nearly all of the codes used once in our sample set have been used again.

As mentioned in the Methods section, we emailed some code authors when we did not find a download site for a package through other methods, sending 46 emails representing 49 different codes. Of the 19 replies we received (by 30 September 2017), 12 expressed an interest in or intent to release their software eventually, with the timeline to making the code public ranging from a few weeks to a few years. One query provided source code to be housed on the ASCL, thus making it available, and we received URLs for software download sites for three codes that our searching had been unable to find. As a result of our email campaign, four codes moved from the \textit{N} category (``we cannot find it") to the \textit{A} category (``source code is available for download without restriction or impediment"), though one would not have been available otherwise.

In a quick evaluation done in 2012, our response rate to similar emails to software authors was 33\%, divided among those who let us know the software we were looking for was not available (20\%) and those that yielded source code for download (13\%), as mentioned in \citet{Shamiretal2013}. We are very pleased to have a response rate of 41\% to our recent inquiries, though our true goal is code yield -- how many codes of those we asked for were or became accessible; this effort gave us a yield of 8.7\% (4 codes/46 emails).

\subsection{Link checking}  \label{subsec:resultslinks}
Our script extracted 2,591 hyperlinks from the PDFs of every research paper published in \textit{A\&A} in 2015, excluding those for the eight domains mentioned previously. After removing malformed links, 2,558 hyperlinks remained. Of these, 30 used the FTP protocol; the rest used HTTP(S). We checked the FTP links by hand and found that four were consistently unreachable, one was intermittently inaccessible, and the other 25 were consistently accessible (see Table \ref{tab:ftplinks}).

\begin{deluxetable}
{c|cc}
\tablecaption{FTP links accessibility status \\
Total links: 30 \label{tab:ftplinks}}
\tablehead{
\colhead{Status} & \colhead{Number of links} \\
}
\startdata
Always accessible & 25 \\
Intermittently accessible & 1 \\
Always inaccessible & 4 \\
\enddata
\end{deluxetable}

We checked the HTTP(S) links with \textit{check\_links.py} four times, on two different computers in two different locations on four different dates in September and October 2017, to avoid counting links as broken that were only temporarily unavailable due to (for example) routine server maintenance. We found that 267 HTTP(S) links and four FTP links, or 10.6\% of all links, were unreachable (did not return a 200 OK) in all runs of the script. See Table \ref{tab:linkfailcodes} for the error codes returned by the unreachable HTTP(S) links.

This statistic, however, differs slightly from the true number of unreachable links. Three links returned 3XX redirect codes, which were counted as unreachable but should not be assumed to be so (in fact, one was accessible), and ten links were deemed inaccessible due to SSL certificate errors, though manual checking showed that nine were accessible. On the other hand, some links that returned inconsistent results likely became inaccessible in the period between the first and last runs of the script. However, most consistently inaccessible links returned status codes such as 404 or -1 (our coding for inability to reach the site at all), which are not likely to be temporary conditions or actually accessible, and there were very few (65) inconsistently accessible links, so these factors have a negligible impact on the end result: if only the 244 HTTP(S) links that consistently returned the status codes -1, 403, and 404 are truly inaccessible (a clear underestimate), 9.5\% of links are inaccessible, and if all 338 links that we failed to access even once are truly inaccessible (a clear overestimate), 13.2\% are inaccessible.

\begin{deluxetable}
{c|cc}
\tablecaption{Error codes for and number of links \\
consistently unreachable \tablenotemark{a} \label{tab:linkfailcodes}}
\tablehead{
\colhead{Error code} & \colhead{Message} & \colhead{Number of} \\
\colhead{} & \colhead{} & \colhead{links}
}
\startdata
-4 & httplib.BadStatusLine & 1 \\
-3 & socket.error & 2 \\
-2 & ssl.CertificateError & 10 \\
-1 & urllib2.URLError (lookup failed) & 74 \\
301 & Moved permanently (redirect) & 2 \\
302 & Found & 1 \\
401 & Unauthorized & 2 \\
403 & Forbidden & 25 \\
404 & Not found & 145 \\
500 & Internal server error & 4 \\
502 & Bad gateway & 0 \\
503 & Service unavailable & 1 \\
504 & Gateway timeout & 0 \\
\enddata
\tablenotetext{a}{Negative error codes denote our coding for links that did not supply an error message}
\end{deluxetable}

\section{Discussion and conclusion}  \label{sec:discussion}
As of 30 September 2017, the codes used in this body of research -- in our 166 articles -- that are immediately available for a reader of these articles are those in the \textit{A} and the \textit{GS} categories, 57.5\% of the software identified in these papers. If we add codes from the \textit{P1} category, for purchasable source codes as in \textit{Numerical Recipes} (and indeed, all codes in this category came from one edition or another of that resource), a volume that is commonly available to researchers, the figure rises to 59.6\%, 170 of the 285 unique codes from our examined articles.

We were unable to find source code online, or it has a hard gate, is available only as an executable or only to collaborators, must be purchased, is a web-accessible black box, or is otherwise not immediately available to the researcher who might like to see how something was calculated, for 115 codes, 40.4\% of the codes used in the research we looked at. 

What does this mean for the repeatability of the research shared in these articles? And how transparent is any individual paper? 

Certainly the relatively broad use of IRAF in research -- appearing in 31 papers (as components or in whole), 18.7\% of our sample -- and other popular open access packages has a positive impact on the transparency of astrophysics research. We would expect to see AstroPy (AstroPy Collaboration, 2013), for example, occupy a similar spot in future research. Such packages improve the discipline in a number of ways in addition to providing transparency, including improving efficiency, as others do not have to write software for the functionality these packages provide. 

Indeed, when we look not at the statistics for unique codes, but instead at the use of software spread throughout the body of these 166 papers, the availability of source code increases in comparison. Taking the use of category \textit{A}, \textit{GS}, and \textit{P1} codes across all the papers, 65.3\% of the software used in the papers had source code available. The use of IRAF and other open source packages such as SExtractor \citep{SExtractor1996}, GILDAS \citep{GILDAS}, and CASA \citep{CASA} in multiple papers weights the overall ``availability index," we might say, favorably upward.

There is no guarantee that the code we found was the version used in the article(s) in which we found the software, as we did not look for any specific version. Our ability to find the software now does not mean the software was available in 2015 when these articles were published, nor does our inability to find source code today in 2017 mean that it was unavailable when the research was published. This study is a snapshot in time; we do not make any claims about our numbers standing for software availability for astrophysics research overall.

At the time we conducted our research (August-October 2017), the articles we examined had been published only two years before. Assuming the ``best case" for the hyperlinks we tested, 9.7\% of the links were inaccessible when we tested them. We have made the links SQLite database available on the aforementioned GitHub repository, and have also made it available in Figshare. We plan to test these links periodically to see what the rate of loss is over time, and we plan to build out the database in the future to get a more complete picture of link persistence. 

It would be interesting to categorize the links into categories by their destinations, whether for software, data, websites, or other resources, and compare link accessibility not only through time but also by destination category, and we may do this in subsequent research.

Our preliminary efforts may be flawed in various ways; we may have missed information about code use in the articles we looked at and included information about data in our larger dataset of 715 entries. In fact, we know this latter happened, but this is mitigated in this study by our excluding records that did not clearly identify software usage. In particular, we might be under-reporting commercial code use slightly, given the first author's proclivity to focus on researcher-written software. We may have miscategorized one or more software packages, either through an inability to find a site for the source code, lack of knowledge or experience with one or more packages and components, or through misunderstanding text about or reference for a code. We may have been overly strict about some software, such as the Lomb-Scargle periodigram algorithm, in wanting to know exactly which implementation was used. In not scoring records that did not have easily identifiable software, we almost certainly left software to examine on the table. It is very likely we did not find all the URLs in the articles we examined, as we pulled back only those that were coded as active hyperlinks, and we may have errors in one or more of our scripts. We have made all of our software, the records used for this paper, and other artifacts of our research available so that others can vet the information we provide here.

Still, this is a beginning, and we are heartened to see more software has source code available than we initially thought and thank the software and article authors for making that so. We will make use of what we have learned in conducting this study and apply it to the next.

\section{Acknowledgments} \label{sec:acknowledgments}
This research has made use of NASA's Astrophysics Data System; indeed, this research would have been nearly impossible without it. Our thanks to Kimberly DuPrie, Lior Shamir, Bruce Berriman, and Robert Nemiroff for reading articles for software usage mentions, to Lior Shamir, Robert Nemiroff, Bruce Berriman, Tracy Huard, James Howison, and Robert Haines for helpful conversations, and to James Howison for continuing email correspondence about this project. The anonymous referee provided encouragement and a helpful report and we are grateful for this input and thank the referee for it. We thank the authors of the papers we examined, the scientists who wrote the software that enabled that research, and \textit{A\&A} for making its articles open access.

\bibliographystyle{aasjournal}
\bibliography{ascl}

\begin{thebibliography}{}
\expandafter\ifx\csname natexlab\endcsname\relax\def\natexlab#1{#1}\fi
\providecommand{\url}[1]{\href{#1}{#1}}

\bibitem[{Ahalt {et~al.}(2015)Ahalt, Carsey, Couch, Hooper, Ibanez, Idaszak,
  Jones, Lin, \& Robinson}]{Ahaltetal2015}
Ahalt, S., Carsey, T., Couch, A., {et~al.} 2015, {NSF Workshop on Supporting
  Scientific Discovery Through Norms and Practices for Software and Data
  Citation and Attribution}, Tech. rep., USA

\bibitem[{{Allen} {et~al.}(2013){Allen}, {Berriman}, {Brunner}, {Burger},
  {DuPrie}, {Hanisch}, {Mann}, {Mink}, {Sandin}, {Shortridge}, \&
  {Teuben}}]{2013ASPC..475..383A}
{Allen}, A., {Berriman}, B., {Brunner}, R., {et~al.} 2013, in Astronomical
  Society of the Pacific Conference Series, Vol. 475, Astronomical Data
  Analysis Software and Systems XXII, ed. D.~N. {Friedel}, 383

\bibitem[{{Baker}(2016)}]{Baker2016}
{Baker}, M. 2016, \nat, 533, 452

\bibitem[{Barnes(2010)}]{barnes2010}
Barnes, N. 2010, Nature, 467, doi:10.1038/467753a.
\newblock \url{http://www.nature.com/news/2010/101013/full/467753a.html}

\bibitem[{{Bertin} \& {Arnouts}(1996)}]{SExtractor1996}
{Bertin}, E., \& {Arnouts}, S. 1996, \aaps, 117, 393

\bibitem[{{Bonanno} \& {Urpin}(2015)}]{BonannoUrpin2015}
{Bonanno}, A., \& {Urpin}, V. 2015, \aap, 574, A63

\bibitem[{Brown {et~al.}(2007)Brown, Brady, Dietz, Cao, Johnson, \&
  McNabb}]{Brownetal2007}
Brown, D.~A., Brady, P.~R., Dietz, A., {et~al.} 2007, {A Case Study on the Use
  of Workflow Technologies for Scientific Analysis: Gravitational Wave Data
  Analysis}, ed. I.~J. Taylor, E.~Deelman, D.~B. Gannon, \& M.~Shields (London:
  Springer London), 39--59.
\newblock \url{https://doi.org/10.1007/978-1-84628-757-2_4}

\bibitem[{{Campana}(2017)}]{Campana2017}
{Campana}, R. 2017, {pyLCSIM: X-ray lightcurves simulator}, Astrophysics Source
  Code Library, , , ascl:1708.016

\bibitem[{Collberg \& Proebsting(2016)}]{CollbergProebsting2016}
Collberg, C., \& Proebsting, T.~A. 2016, Commun. ACM, 59, 62.
\newblock \url{http://doi.acm.org/10.1145/2812803}

\bibitem[{{Fenniak}(2014)}]{PYPDF}
{Fenniak}, M. 2014, {pyPdf: PDF Toolkit}, Python Package Index,  Python Package
  Index.
\newblock \url{https://pypi.python.org/pypi/pyPdf}

\bibitem[{{GILDAS Team}(2013)}]{GILDAS}
{GILDAS Team}. 2013, {GILDAS: Grenoble Image and Line Data Analysis Software},
  Astrophysics Source Code Library, , , ascl:1305.010

\bibitem[{Goble {et~al.}(2016)Goble, Howison, Kirchner, Nierstrasz, \&
  Vinju}]{Gobleetal2016}
Goble, C., Howison, J., Kirchner, C., Nierstrasz, O., \& Vinju, J.~J. 2016,
  Dagstuhl Reports, 6, 62.
\newblock \url{http://drops.dagstuhl.de/opus/volltexte/2016/6755}

\bibitem[{Hettrick {et~al.}(2014)Hettrick, Antonioletti, Carr, Chue~Hong,
  Crouch, De~Roure, Emsley, Goble, Hay, Inupakutika, Jackson, Nenadic,
  Parkinson, Parsons, Pawlik, Peru, Proeme, Robinson, \& Sufi}]{hettrick2014}
Hettrick, S., Antonioletti, M., Carr, L., {et~al.} 2014, {UK Research Software
  Survey 2014}, , , doi:10.5281/zenodo.14809.
\newblock \url{https://doi.org/10.5281/zenodo.14809}

\bibitem[{Howison \& Bullard(2016)}]{HowisonBullard2016}
Howison, J., \& Bullard, J. 2016, Journal of the Association for Information
  Science and Technology, 67, 2137.
\newblock \url{http://dx.doi.org/10.1002/asi.23538}

\bibitem[{Howison {et~al.}(2015)Howison, Deelman, McLennan, Ferreira~da Silva,
  \& Herbsleb}]{Howisonetal2015}
Howison, J., Deelman, E., McLennan, M.~J., Ferreira~da Silva, R., \& Herbsleb,
  J.~D. 2015, Research Evaluation, 24, 454.
\newblock \url{+ http://dx.doi.org/10.1093/reseval/rvv014}

\bibitem[{{Ince} {et~al.}(2012){Ince}, {Hatton}, \&
  {Graham-Cumming}}]{Inceetal2012}
{Ince}, D.~C., {Hatton}, L., \& {Graham-Cumming}, J. 2012, \nat, 482, 485

\bibitem[{{Joppa} {et~al.}(2013){Joppa}, {McInerny}, {Harper}, {Salido},
  {Takeda}, {O'Hara}, {Gavaghan}, \& {Emmott}}]{Joppaetal2013}
{Joppa}, L.~N., {McInerny}, G., {Harper}, R., {et~al.} 2013, Science, 340, 814

\bibitem[{{Kalamkar} {et~al.}(2016){Kalamkar}, {Casella}, {Uttley}, {O'Brien},
  {Russell}, {Maccarone}, {van der Klis}, \& {Vincentelli}}]{Kalamkaretal2016}
{Kalamkar}, M., {Casella}, P., {Uttley}, P., {et~al.} 2016, \mnras, 460, 3284

\bibitem[{{Katz} {et~al.}(2014){Katz}, {Choi}, {Lapp}, {Maheshwari},
  {L{\"o}ffler}, {Turk}, {Hanwell}, {Wilkins-Diehr}, {Hetherington}, {Howison},
  {Swenson}, {Allen}, {Elster}, {Berriman}, \& {Venters}}]{Katzetal2014}
{Katz}, D.~S., {Choi}, S.-C.~T., {Lapp}, H., {et~al.} 2014, ArXiv e-prints,
  arXiv:1404.7414

\bibitem[{Limare(2012)}]{limare2012}
Limare, N. 2012, in {ICERM Workshop on Reproducibility in Computational and
  Experimental Mathematics}, Providence, United States,
  http://icerm.brown.edu/tw12-5-rcem.
\newblock \url{https://hal.archives-ouvertes.fr/hal-00783292}

\bibitem[{{Lomb}(1976)}]{Lomb1976}
{Lomb}, N.~R. 1976, \apss, 39, 447

\bibitem[{{Ludaescher} {et~al.}(2016){Ludaescher}, {Chard}, {Gaffney}, {Jones},
  {Nabrzyski}, {Stodden}, \& {Turk}}]{Ludaescheretal1016}
{Ludaescher}, B., {Chard}, K., {Gaffney}, N., {et~al.} 2016, ArXiv e-prints,
  arXiv:1610.09958

\bibitem[{Marwick(2015)}]{Marwick2015}
Marwick, B. 2015, {How computers broke science \&ndash; and what we can do to
  fix it}, , .
\newblock
  \url{http://theconversation.com/how-computers-broke-science-and-what-we-can-do-to-fix-it-49938}

\bibitem[{{McMullin} {et~al.}(2007){McMullin}, {Waters}, {Schiebel}, {Young},
  \& {Golap}}]{CASA}
{McMullin}, J.~P., {Waters}, B., {Schiebel}, D., {Young}, W., \& {Golap}, K.
  2007, in Astronomical Society of the Pacific Conference Series, Vol. 376,
  Astronomical Data Analysis Software and Systems XVI, ed. R.~A. {Shaw},
  F.~{Hill}, \& D.~J. {Bell}, 127

\bibitem[{{Momcheva} \& {Tollerud}(2015)}]{MomchevaTollerud2015}
{Momcheva}, I., \& {Tollerud}, E. 2015, ArXiv e-prints, arXiv:1507.03989

\bibitem[{{Morin} {et~al.}(2012){Morin}, {Urban}, {Adams}, {Foster}, {Sali},
  {Baker}, \& {Sliz}}]{Morinetal2012}
{Morin}, A., {Urban}, J., {Adams}, P.~D., {et~al.} 2012, Science, 336, 159

\bibitem[{NASA(2014)}]{nasadatacitation}
NASA. 2014, {NASA Plan for Increasing Access to the Results of Scientific
  Research}, Tech. rep.
\newblock
  \url{http://www.nasa.gov/sites/default/files/atoms/files/206985_2015_nasa_plan-for-web.pdf}

\bibitem[{{NASA High Energy Astrophysics Science Archive Research Center
  (Heasarc)}(2014)}]{HEASOFT}
{NASA High Energy Astrophysics Science Archive Research Center (Heasarc)}.
  2014, {HEAsoft: Unified Release of FTOOLS and XANADU}, Astrophysics Source
  Code Library, , , ascl:1408.004

\bibitem[{{Press} {et~al.}(1986){Press}, {Flannery}, \&
  {Teukolsky}}]{Press1986}
{Press}, W.~H., {Flannery}, B.~P., \& {Teukolsky}, S.~A. 1986, {Numerical
  recipes. The art of scientific computing}

\bibitem[{{Press} {et~al.}(1992{\natexlab{a}}){Press}, {Teukolsky},
  {Vetterling}, \& {Flannery}}]{Press1992C}
{Press}, W.~H., {Teukolsky}, S.~A., {Vetterling}, W.~T., \& {Flannery}, B.~P.
  1992{\natexlab{a}}, {Numerical recipes in C. The art of scientific computing}

\bibitem[{{Press} {et~al.}(1992{\natexlab{b}}){Press}, {Teukolsky},
  {Vetterling}, \& {Flannery}}]{Press1992F}
---. 1992{\natexlab{b}}, {Numerical recipes in FORTRAN. The art of scientific
  computing}

\bibitem[{{Press} {et~al.}(2002){Press}, {Teukolsky}, {Vetterling}, \&
  {Flannery}}]{Press2002C}
---. 2002, {Numerical recipes: the art of scientific computing}

\bibitem[{{Santos-Sanz} {et~al.}(2015){Santos-Sanz}, {Ortiz}, {Morales},
  {Duffard}, {Pozuelos}, {Moreno}, \&
  {Fern{\'a}ndez-Valenzuela}}]{Santos-Sanzetal2015}
{Santos-Sanz}, P., {Ortiz}, J.~L., {Morales}, N., {et~al.} 2015, \aap, 575, A52

\bibitem[{{Scargle}(1982)}]{Scargle1982}
{Scargle}, J.~D. 1982, \apj, 263, 835

\bibitem[{{Shamir} {et~al.}(2013){Shamir}, {Wallin}, {Allen}, {Berriman},
  {Teuben}, {Nemiroff}, {Mink}, {Hanisch}, \& {DuPrie}}]{Shamiretal2013}
{Shamir}, L., {Wallin}, J.~F., {Allen}, A., {et~al.} 2013, Astronomy and
  Computing, 1, 54

\bibitem[{{Smith} {et~al.}(2016){Smith}, {Katz}, {Niemeyer}, \& {FORCE11
  Software Citation Working Group}}]{softwarecitationprinciples}
{Smith}, A.~M., {Katz}, D.~S., {Niemeyer}, K.~D., \& {FORCE11 Software Citation
  Working Group}. 2016, PeerJ Computer Science, 2:e86, doi:10.7717/peerj-cs.86

\bibitem[{{Smith} {et~al.}(2017){Smith}, {E Niemeyer}, {Katz}, {Barba},
  {Githinji}, {Gymrek}, {Huff}, {Madan}, {Cabunoc Mayes}, {Moerman}, {Prins},
  {Ram}, {Rokem}, {Teal}, {Valls Guimera}, \& {Vanderplas}}]{Smithetal2017}
{Smith}, A.~M., {E Niemeyer}, K., {Katz}, D.~S., {et~al.} 2017, ArXiv e-prints,
  arXiv:1707.02264

\bibitem[{Soito \& Hwang(2016)}]{Soito_2016}
Soito, L., \& Hwang, L.~J. 2016, International Journal of Digital Curation, 11,
  doi:10.2218/ijdc.v11i2.390.
\newblock \url{https://doi.org/10.2218%2Fijdc.v11i2.390}

\bibitem[{{Stetson}(1987)}]{DAOPHOT}
{Stetson}, P.~B. 1987, \pasp, 99, 191

\bibitem[{Stodden(2010)}]{Stodden2010}
Stodden, V. 2010, {The Scientific Method in Practice: Reproducibility in the
  Computational Sciences}, Tech. Rep. MIT Sloan Research Paper No. 4773-10,
  doi:10.2139/ssrn.1550193

\bibitem[{{Stodden} {et~al.}(2016){Stodden}, {McNutt}, {Bailey}, {Deelman},
  {Gil}, {Hanson}, {Heroux}, {Ioannidis}, \& {Taufer}}]{Stoddenetal2016}
{Stodden}, V., {McNutt}, M., {Bailey}, D.~H., {et~al.} 2016, Science, 354, 1240

\bibitem[{{Taddia} {et~al.}(2015){Taddia}, {Sollerman}, {Fremling},
  {Pastorello}, {Leloudas}, {Fransson}, {Nyholm}, {Stritzinger}, {Ergon},
  {Roy}, \& {Migotto}}]{Taddiaetal2015}
{Taddia}, F., {Sollerman}, J., {Fremling}, C., {et~al.} 2015, \aap, 580, A131

\bibitem[{{Teuben} {et~al.}(2014){Teuben}, {Allen}, {Berriman}, {DuPrie},
  {Hanisch}, {Mink}, {Nemiroff}, {Shamir}, {Shortridge}, {Taylor}, \&
  {Wallin}}]{Teubenetal2014}
{Teuben}, P., {Allen}, A., {Berriman}, B., {et~al.} 2014, in Astronomical
  Society of the Pacific Conference Series, Vol. 485, Astronomical Data
  Analysis Software and Systems XXIII, ed. N.~{Manset} \& P.~{Forshay}, 3

\bibitem[{{Tody}(1986)}]{IRAF1}
{Tody}, D. 1986, in \procspie, Vol. 627, Instrumentation in astronomy VI, ed.
  D.~L. {Crawford}, 733

\bibitem[{{Tody}(1993)}]{IRAF2}
{Tody}, D. 1993, in Astronomical Society of the Pacific Conference Series,
  Vol.~52, Astronomical Data Analysis Software and Systems II, ed. R.~J.
  {Hanisch}, R.~J.~V. {Brissenden}, \& J.~{Barnes}, 173

\bibitem[{{Tubiana} {et~al.}(2015){Tubiana}, {Snodgrass}, {Bertini}, {Mottola},
  {Vincent}, {Lara}, {Fornasier}, {Knollenberg}, {Thomas}, {Fulle}, {Agarwal},
  {Bodewits}, {Ferri}, {G{\"u}ttler}, {Gutierrez}, {La Forgia}, {Lowry},
  {Magrin}, {Oklay}, {Pajola}, {Rodrigo}, {Sierks}, {A'Hearn}, {Angrilli},
  {Barbieri}, {Barucci}, {Bertaux}, {Cremonese}, {Da Deppo}, {Davidsson}, {De
  Cecco}, {Debei}, {Groussin}, {Hviid}, {Ip}, {Jorda}, {Keller}, {Koschny},
  {Kramm}, {K{\"u}hrt}, {K{\"u}ppers}, {Lazzarin}, {Lamy}, {Lopez Moreno},
  {Marzari}, {Michalik}, {Naletto}, {Rickman}, {Sabau}, \&
  {Wenzel}}]{Tubianaetal2015}
{Tubiana}, C., {Snodgrass}, C., {Bertini}, I., {et~al.} 2015, \aap, 573, A62

\bibitem[{{VanderPlas}(2017)}]{VanderPlas2017}
{VanderPlas}, J.~T. 2017, ArXiv e-prints, arXiv:1703.09824

\bibitem[{Vitek \& Kalibera(2011)}]{VitekKalibera2011}
Vitek, J., \& Kalibera, T. 2011, in Proceedings of the Ninth ACM International
  Conference on Embedded Software, EMSOFT '11 (New York, NY, USA: ACM), 33--38.
\newblock \url{http://doi.acm.org/10.1145/2038642.2038650}

\bibitem[{{Vrard} {et~al.}(2015){Vrard}, {Mosser}, {Barban}, {Belkacem},
  {Elsworth}, {Kallinger}, {Hekker}, {Samadi}, \& {Beck}}]{Vrardetal2015}
{Vrard}, M., {Mosser}, B., {Barban}, C., {et~al.} 2015, \aap, 579, A84

\bibitem[{{Weiner} {et~al.}(2009){Weiner}, {Blanton}, {Coil}, {Cooper},
  {Dav{\'e}}, {Hogg}, {Holden}, {Jonsson}, {Kassin}, {Lotz}, {Moustakas},
  {Newman}, {Prochaska}, {Teuben}, {Tremonti}, \& {Willmer}}]{Weineretal2009}
{Weiner}, B., {Blanton}, M.~R., {Coil}, A.~L., {et~al.} 2009, in ArXiv
  Astrophysics e-prints, Vol. 2010, astro2010: The Astronomy and Astrophysics
  Decadal Survey, 61P

\end{thebibliography}

\appendix
\section{Code frequency and category} \label{app:A}

\startlongtable
\begin{deluxetable}{p{6cm}|cc}

\tablewidth{0pt}
\tabletypesize{\small}
\tablecaption{Code frequency and category}

\tablehead{
\colhead{Code name/text} & \colhead{Frequency} & \colhead{Category} \\
}
\startdata
IRAF & 31 & A \\
SExtractor & 10 & A \\
HIPE & 7 & B \\
GILDAS & 5 & A \\
BGM: Besan\c{c}on Galaxy model & 4 & W \\
CASA & 4 & A \\
CIAO & 4 & A \\
DAOPHOT & 4 & A \\
DAOSPEC & 4 & GH \\
MOOG & 4 & A \\
MPFIT & 4 & A \\
SAS: Science Analysis Software & 4 & GS \\ 
XSPEC & 4 & A \\
ARES & 3 & A \\
BEC & 3 & N \\
ESO GIRAFFE pipeline & 3 & A \\
GARSTEC & 3 & GH \\
Gasgano & 3 & B \\
HEASoft & 3 & A \\
Libre-Esprit & 3 & N \\
MARCS & 3 & A \\
PHOEBE & 3 & A \\
pPXF & 3 & A \\
Reflex & 3 & B \\
Scanamorphos & 3 & A \\
SPICE & 3 & A \\
TOPCAT & 3 & A \\
\enddata
\tablecomments{Table 9 is not published in its entirety here; it will be available in a machine-readable format when the article is published. A portion is shown here for guidance regarding its form and content.}

\end{deluxetable}

\newpage
\section{Sample email to software authors} \label{app:B}

\begin{verbatim}
Subject line: [code name] 

Dear Dr. [name], 

I came across [code name] when looking for astrophysics codes. I would
like to add this code to the Astrophysics Source Code Library (ASCL,
ascl.net). The ASCL is indexed by ADS and Web of Science; this
indexing provides useful research software with greater visibility.

I could not find a link to the source code, however. Is the code
available for download, and if so, what is the URL? If there is no
download site, the ASCL can house an archive file of the code if you
prefer.

The ASCL seeks to increase the transparency of astrophysics research
by making codes discoverable for examination and we are always
looking for codes which have been used in refereed research. If you
have additional codes that should be listed, I am interested in
knowing about them.

Thank you for your attention. 

Clear skies, 

Alice Allen 
Editor, ASCL
\end{verbatim}
\newpage
\section{IDs and links to PDFs of examined articles} \label{app:C}

\begin{deluxetable}
{c|c}
\tablecaption{IDs and links to PDFs of examined articles \\}
\tablehead{
\colhead{Article ID} & \colhead{Link to PDF} \\
}
\startdata
aa23704-14 & https://www.aanda.org/articles/aa/pdf/2015/01/aa23704-14.pdf \\
aa23996-14 & https://www.aanda.org/articles/aa/pdf/2015/01/aa23996-14.pdf \\
aa24235-14 & https://www.aanda.org/articles/aa/pdf/2015/01/aa24235-14.pdf \\
aa24356-14 & https://www.aanda.org/articles/aa/pdf/2015/01/aa24356-14.pdf \\
aa24393-14 & https://www.aanda.org/articles/aa/pdf/2015/01/aa24393-14.pdf \\
aa24472-14 & https://www.aanda.org/articles/aa/pdf/2015/01/aa24472-14.pdf \\
aa24589-14 & https://www.aanda.org/articles/aa/pdf/2015/01/aa24589-14.pdf \\
aa24675-14 & https://www.aanda.org/articles/aa/pdf/2015/01/aa24675-14.pdf \\
aa24691-14 & https://www.aanda.org/articles/aa/pdf/2015/01/aa24691-14.pdf \\
aa24735-14 & https://www.aanda.org/articles/aa/pdf/2015/01/aa24735-14.pdf \\
aa24967-14 & https://www.aanda.org/articles/aa/pdf/2015/01/aa24967-14.pdf \\
\enddata
\tablecomments{Table 10 is not published in its entirety here; it will be available in a machine-readable format when the article is published. A portion is shown here for guidance regarding its form and content.}

\end{deluxetable}

\end{document}